\documentclass[pra, twocolumn, showpacs,
superscriptaddress , unsortedaddress, preprintnumbers]{revtex4}

\usepackage{amsmath}

\usepackage{graphicx}

\renewcommand{\Re}{\mathop{\mathrm{Re}}\nolimits}
\newcommand{\D}{\mbox{\rm d}}
\newcommand{\T}{\mathcal{T}}
\newcommand{\R}{\mathcal{R}}
\newcommand{\A}{\mathcal{A}}

\begin{document}
\preprint{PHYSICAL REVIEW A {\bf 75}, 013807 (2007)}
\title{Determination of quantum-noise parameters of realistic cavities}
\author{A.A. Semenov}
\email[E-mail address: ]{sem@iop.kiev.ua} \affiliation{
Institut f\"ur Physik, Universit\"{a}t Rostock, Universit\"{a}tsplatz
3, D-18051 Rostock, Germany} \affiliation{Institute of Physics,
National Academy of Sciences of Ukraine, Prospect Nauky 46,
UA-03028 Kiev, Ukraine}

\author{W. Vogel}
\affiliation{ Institut f\"ur Physik, Universit\"{a}t Rostock,
Universit\"{a}tsplatz 3, D-18051 Rostock, Germany}

\author{M. Khanbekyan}
\affiliation{Theoretisch-Physikalisches Institut,
Friedrich-Schiller-Universit\"{a}t Jena, Max-Wien-Platz 1, D-07743
Jena, Germany}

\author{D.-G. Welsch}
\affiliation{Theoretisch-Physikalisches Institut,
Friedrich-Schiller-Universit\"{a}t Jena, Max-Wien-Platz 1, D-07743
Jena, Germany}
\date{\today}
\begin{abstract}
  A procedure is developed which allows one to measure all the parameters
  occurring in a complete model
  [A.A. Semenov et al., Phys. Rev. A {\bf 74}, 033803 (2006); quant-ph/0603043] of realistic leaky cavities with unwanted noise. The method
  is based on the reflection of properly chosen test pulses by the cavity.
\end{abstract}

\maketitle

\section{Introduction}

Optical cavities play an important role for a variety of experiments
in quantum optics~\cite{CQED}. The possibilities of realizing strong
coupling between cavity modes and atoms inside the cavities make
them of interest for the efficient transfer of quantum states
between the radiation field
and atoms, as desired in quantum information processing. 
For describing the coupling of radiation into and out of a cavity, quantum
stochastic input-output relations~\cite{Collett} and quantum field theoretical
concepts were developed~\cite{Knoll}.

A realistic description of the input-output behavior of leaky
cavities requires a careful consideration of possible losses that
may significantly alter the nonclassical properties of light. Cavity
parameters, such as the transmission and loss coefficients, the free
spectral range (FSR), the cavity-decay rate $\Gamma$, and the
$Q$-factor, have been measured by using the direct transmission of
light through the cavity \cite{Hood,Trope} and by monitoring
nonclassical charcteristics of light \cite{Mikhailov}. It has been
demonstrated that in high-$Q$ cavities the losses caused by
absorption and scattering may be of the same order of magnitude as
the wanted outcoupling of the field~\cite{Hood}. For describing
additional effects of unwanted losses associated with absorption and
scattering, attempts have been made to introduce additional
input-output ports into the quantum noise theory, see e.g.
Refs.~\cite{Khanbekyan, Viviescas}. It has been shown that even
small unwanted losses due to absorption and scattering may
substantially diminish nonclassical signatures of the outgoing
pulse~\cite{Khanbekyan}. Hence for any application that requires a
precise transfer of quantum states of light into or out of a cavity,
a careful description of the properties of the used cavity is
indispensable.

Recently it has been demonstrated that a complete cavity model,
describing all the unwanted losses, requires additional noise terms
in both the quantum Langevin equation and the input-output
relation~\cite{Semenov}. This may lead to new effects, such as the
superposition of the input field with the outgoing field from the
cavity in a common nonmonochromatic mode. In principle, all
parameters necessary for the complete description of cavities can be
expressed in terms of the radiation and absorption coefficients of
the mirrors~\cite{Khanbekyan2}. In practice, however, the parameters
used in such a model will also depend on scattering, mirror
alignment, and other conditions of the experiment. Hence, an
operational procedure for the determination of all the cavity
parameters is desired for a correct descriptions of the quantum
effects to be expected.

In the present contribution we propose a method to determine the
cavity parameters needed for a complete description of the quantum
noise effects of a realistic cavity. It is based on the reflection
of light pulses of different lengths by the cavity. The absorption
and scattering of pulses with a spectrum that is much wider than the
cavity decay rate is shown to depend only on the additional term in
the input-output relation. This fact will appear to be useful for
the needed measurement procedure.

\section{The cavity model}

Let us consider the recently proposed model of leaky cavity, which
has a partial transmitting mirror, with unwanted
noise~\cite{Semenov}. The cavity is described by the quantum
Langevin equation
\begin{equation}
\label{QLE}
\dot{\hat{a}}_\mathrm{cav}=-\left[i\omega_\mathrm{cav}+
{\textstyle \frac{1}{2} }
\Gamma\right]\hat{a}_\mathrm{cav}
 +\,\T^\mathrm{(c)}\hat{b}_\mathrm{in}\left(t_1\right)
+\hat{C}^{(c)}\left(t_1\right),
\end{equation}
and the input-output relation
\begin{equation}
\label{IOR}
\hat{b}_\mathrm{out}\!\left(t_1\right)=
\T^\mathrm{(o)}\hat{a}_\mathrm{cav}\!\left(t_1\right)
+\R^\mathrm{(o)}\hat{b}_\mathrm{in}\!\left(t_1\right)
+\hat{C}^{(o)}\!\left(t_1\right).
\end{equation}
Here, $\hat{a}_\mathrm{cav}$ is the annihilation operator of an intracavity
mode, $\hat{b}_\mathrm{in}\!\left(t_1\right)$ is the input-field operator,
$\hat{b}_\mathrm{out}\!\left(t_1\right)$ is the output-field operator,
$\omega_\mathrm{cav}$ is the resonance frequency of the cavity, and $\Gamma$
is the cavity decay rate. The complex numbers $\T^\mathrm{(o)}$ and
$\T^\mathrm{(c)}$ are the transmission coefficients describing the outcoupling
of the internal field and the incoupling of the input field respectively.  The
complex number $\R^\mathrm{(o)}$ is the reflection coefficient at the cavity.
The operators $\hat{C}^{(c)}\!\left(t_1\right)$ and
$\hat{C}^{(o)}\!\left(t_1\right)$ are associated with the unwanted losses and
obey the commutation relations
\begin{align}
\label{cr1}
&\bigl[\hat{C}^{(c)}(t_1),\hat{C}^{(c)\dag}(t_2)\bigr]=
\bigl|\A^{(c)}\bigr|^2\delta(t_1-t_2),
\\
\label{cr2}
&\bigl[\hat{C}^{(o)}(t_1),\hat{C}^{(o)\dag}(t_2)\bigr]=
\bigl|\A^{(o)}\bigr|^2\delta(t_1-t_2),
\\
\label{cr3} &\bigl[\hat{C}^{(c)}(t_1),\hat{C}^{(o)\dag}(t_2)\bigr]=
\bigl|\A^{(c)}\bigr|\bigl|\A^{(o)}\bigr|e^{i\kappa}\cos\zeta\,\delta(t_1-t_2)
,
\end{align}
where the numbers $\bigl|\A^{(c)}\bigr|$, $\bigl|\A^{(o)}\bigr|$,
and $e^{i\kappa}\cos\zeta$ are related to the transmission and
reflection coefficients through the constraints
\begin{equation}
\Gamma=\bigl|{\A}^\mathrm{(c)}\bigr|^2+
\bigl|\T^\mathrm{(c)}\bigr|^2,
\label{constr1}
\end{equation}
\begin{equation}
\bigl|\R^\mathrm{(o)}\bigr|^2+\bigl|\A^\mathrm{(o)}\bigr|^2=1, \label{constr2}
\end{equation}
\begin{equation}
\T^\mathrm{(o)}+\T^{\mathrm{(c)}\ast}\R^\mathrm{(o)}+
\bigl|\A^{(c)}\bigr|\bigl|\A^{(o)}\bigr|e^{i\kappa}\cos\zeta=0,\label{constr3}
\end{equation}
which follow from the requirements of preserving the commutation
rules, for details see Ref.~\cite{Semenov}.

The set of complex coefficients $\omega_\mathrm{cav}$, $\Gamma$,
$\T^\mathrm{(o)}$, $\T^\mathrm{(c)}$, and $\R^\mathrm{(o)}$
completely describe the radiative and the noise properties related
to the considered intracavity mode. In view of
Eqs.~(\ref{constr1}-\ref{constr3}), these parameters attain only
values within a restricted domain~\cite{Semenov}.  By supposing that
the resonance frequency $\omega_\mathrm{cav}$ and the FSR are
already known, we will formulate an operational procedure for the
measurement of the other parameters of this set. Our procedure is
based on measuring the reflection efficiencies of different test
pulses by the cavity.

\section{Reflection of quantum light by the cavity}

\begin{figure}[htb!]
\includegraphics[clip=, width=0.82\linewidth]{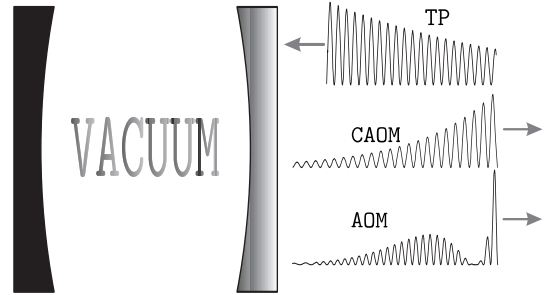}
\caption{\label{fig1} Reflection of a test pulse (TP) by the cavity.
The outgoing field consists of two different nonmonochromatic modes:
the cavity-associated output mode (CAOM) and an additional output
mode (AOM).}
\end{figure}

We start with combining solution of the quantum Langevin equation
(\ref{QLE}) and the input-output relation (\ref{IOR}), also c.f.
\cite{Semenov},
\begin{eqnarray}
&&\hat{b}_\mathrm{out}\left(t_1\right)=\label{IORTime}\\
&&\hat{a}_\mathrm{cav}\!(0)F^\ast\left(t_1\right)+\int_{-\infty}^{+\infty}\!\D
t_2\,
G^\ast\left(t_1,t_2\right)\hat{b}_\mathrm{in}\left(t_2\right)
+\hat{C}\left(t_1\right),\nonumber
\end{eqnarray}
where $\hat{a}_\mathrm{cav}\!(0)$ is the operator of the cavity mode at initial
time,
\begin{equation}
F^\ast\left(t_1\right)=
\T^\mathrm{(o)}e^{-\left(i\omega_\mathrm{cav}+\frac{\Gamma}{2}\right)t_1}
\Theta\left(t_1\right), \label{FTime}
\end{equation}
 \begin{equation}
G^\ast\left(t_1,t_2\right)=\T^\mathrm{(c)}
\xi^\ast\left(t_1,t_2\right)+\R^\mathrm{(o)}\delta
\left(t_1-t_2\right), \label{GTime}
\end{equation}
\begin{equation}
\xi^\ast\left(t_1,t_2\right)= \T^\mathrm{(o)}
e^{-\left(i\omega_\mathrm{cav}+\frac{\Gamma}{2}\right)\left(t_1-t_2\right)}
\Theta\left(t_1\right)\Theta\left(t_1-t_2\right).\label{XiTime}
\end{equation}
Equation (\ref{IORTime}) can be rewritten by using two complete
orthogonal sets of functions
$\left\{U_n^\mathrm{in}\left(t_1\right), n=0,\ldots,+\infty\right\}$
and $\left\{U_n^\mathrm{out}\left(t_1\right),
n=0,\ldots,+\infty\right\}$ associated with input and output fields
respectively,
\begin{equation}
\hat{b}_\mathrm{in(out)}\left(t_1\right)=\sum\limits_{n=0}^{+\infty}
U_n^\mathrm{in(out)}\left(t_1\right)\hat{b}_{\mathrm{in(out)};
n},\label{BDiscrete1}
\end{equation}
\begin{equation}
\hat{b}_{\mathrm{in(out)};n}=\int_{-\infty}^{+\infty} \D t_1
U_n^{\mathrm{in(out)}\ast}\!\left(t_1\right)
\hat{b}_\mathrm{in(out)}\!\left(t_1\right).\label{BDiscrete2}
\end{equation}
Suppose that the function $U_0^\mathrm{out}\!\left(t_1\right)$ has
the form of the pulse extracted from the cavity, i.e., the
cavity-associated output mode (CAOM),
\begin{equation}
U_0^\mathrm{out}\!\left(t_1\right)=\frac{\sqrt{\Gamma}}
{\left|\T^{(o)}\right|}F^\ast\left(t_1\right).\label{U0Out}
\end{equation}
The function $U_0^\mathrm{in}\!\left(t_1\right)$ describes the test pulse (TP)
to be reflected by the cavity, leading to the output pulse
\begin{equation}
U^\mathrm{out}\!\left(t_1\right)=\int_{-\infty}^{+\infty}\D t_2\, G^{\ast}\!
\left(t_1,t_2\right)U_0^\mathrm{in}\!\left(t_2\right).\label{MIMResponse}
\end{equation}
The function $U_1^\mathrm{out}\!\left(t_1\right)$ is chosen such that the
expansion of $U^\mathrm{out}\!\left(t_1\right)$ in terms of
$U_n^\mathrm{out}\!\left(t_1\right)$ consists of two components,
\begin{equation}
U^\mathrm{out}\!\left(t_1\right)=\sqrt{\underline\eta_\mathrm{\,
ref}}e^{i\mu} U_0^\mathrm{out}\!\left(t_1\right)+
\sqrt{\overline\eta_\mathrm{ref}}
U_1^\mathrm{out}\!\left(t_1\right),\label{MIMResponseExp}
\end{equation}
see Fig~\ref{fig1}.

We will refer to the pulse associated with the function
$U_1^\mathrm{out}\!\left(t_1\right)$ as the additional output mode
(AOM). Here $\underline\eta_\mathrm{\, ref}$ and $\mu$ are,
respectively, the efficiency and the phase of the reflection of the
TP into the CAOM.  From the orthogonality conditions we obtain
\begin{eqnarray}
\sqrt{\underline\eta_\mathrm{\, ref}}
e^{i\mu}&=&\int_{-\infty}^{+\infty} \D t_1
U^\mathrm{out}\!\left(t_1\right)
U_0^{\mathrm{out}\ast}\!\left(t_1\right)\label{Eff1}\\
&=&\left[\frac{\T^\mathrm{(o)}\T^\mathrm{(c)}}{\Gamma}+
\R^\mathrm{(o)}\right]\nonumber\\
&\times&\frac{\T^\mathrm{(o)\ast}}{\left|\T^{(o)}\right|}\sqrt{\Gamma}
\int_{-\infty}^{+\infty} \D t_1 U_0^\mathrm{in}\!\left(t_1\right)
e^{(i\omega_\textrm{cav}+\frac{\Gamma}{2})t_1},\nonumber
\end{eqnarray}
where $\overline\eta_\mathrm{ref}$ is the efficiency of the
input-field reflection into the AOM. Using
Eq.~(\ref{MIMResponseExp}) and the normalization condition for
$U_1^\mathrm{out}\!\left(t_1\right)$ one derives
\begin{equation}
\overline\eta_\mathrm{ref}=\int_{-\infty}^{+\infty} \D t_1
\left|U^\mathrm{out}\!\left(t_1\right)-
\sqrt{\underline\eta_\mathrm{\, ref}}e^{i\mu}
U_0^\mathrm{out}\!\left(t_1\right)\right|^2 .\label{Eff2}
\end{equation}
The total efficiency of the input-field reflection at the cavity is given by
the expression
\begin{equation}
\eta_\mathrm{ref}=\underline\eta_\mathrm{\, ref}+\overline\eta_\mathrm{ref}=\int_{-\infty}^{+\infty}
\D t_1 \left|U^\mathrm{out}\!\left(t_1\right)\right|^2 .\label{EtaTotal}
\end{equation}

In the considered representation, Eq.~(\ref{IORTime}) has the form
\begin{eqnarray}
\hat{b}_{\mathrm{out};0}=
\sqrt{\eta_\mathrm{ext}}\,\hat{a}_\mathrm{cav}\!\left(0\right)+
\sqrt{\underline\eta_\mathrm{\, ref}}e^{i\mu}\hat{b}_{\mathrm{in};0}
\nonumber\\+\sum\limits_{m=1}^{+\infty}\!
G^\ast_{m,0}\hat{b}_{\mathrm{in};m}+\hat{C}_0,\label{IORDiscrete0}
\end{eqnarray}
\begin{equation}
\hat{b}_{\mathrm{out};1}=
\sqrt{\overline{\eta}_\mathrm{ref}}\,\hat{b}_{\mathrm{in};0}+
\sum\limits_{m=1}^{+\infty}\! G^\ast_{m,1}\hat{b}_{\mathrm{in};m}
+\hat{C}_{1},\label{IORDiscrete1}
\end{equation}
\begin{equation}
\hat{b}_{\mathrm{out};n}= \sum\limits_{m=1}^{+\infty}\!
G^\ast_{m,n}\hat{b}_{\mathrm{in};m} +\hat{C}_{n}\, \textrm{for
$n=2,3\ldots$}, \label{IORDiscrete2}
\end{equation}
where
\begin{equation}
G^\ast_{m,n}= \int_{-\infty}^{+\infty}\D t_1 \D t_2
U_{n}^{\mathrm{out}\ast}\!\left(t_1\right)G^\ast\left(t_1,t_2\right)
U_{m}^\mathrm{in}\left(t_2\right),\label{GDiscrete}
\end{equation}
\begin{equation}
\hat{C}_{n}=\int_{-\infty}^{+\infty}\D t_1
U_{n}^{\mathrm{out}\ast}\! \left(t_1\right)
\hat{C}\left(t_1\right),\label{CDiscrete}
\end{equation}
\begin{equation}
\eta_\mathrm{ext}=\frac{\left|\T^{(o)}\right|^2}{\Gamma}.\label{EtaExt}
\end{equation}
Equation~(\ref{EtaExt}) defines the efficiency of quantum-state
extraction from the cavity \cite{Khanbekyan}.

Consider the reflection of the TP, whose form is that of a
pulse extracted from another cavity with the same resonance frequency
and the cavity decay rate $\Gamma^{(k)}$,
\begin{equation}
U_0^\mathrm{in}\!\left(t_1\right)=\sqrt{\Gamma^{(k)}}e^{-\left(i\omega_\mathrm{cav}
+\frac{\Gamma^{(k)}}{2}\right)t_1}e^{i\varphi^{(k)}}\Theta\!\left(t_1\right).\label{TestPulse}
\end{equation}
In this case the reflected pulse according to Eq.~(\ref{MIMResponse}) has the form
\begin{eqnarray}
&&U^\mathrm{out}\!\left(t_1\right)=\sqrt{\Gamma^{(k)}}e^{-\left(i\omega_\mathrm{cav}
+\frac{\Gamma^{(k)}}{2}\right)t_1}e^{i\varphi^{(k)}}\Theta\!\left(t_1\right)\label{OutPulse}
\\&&\times \left\{\frac{2\T^{(o)}\T^{(c)}}{\Gamma-\Gamma^{(k)}}
\left[1-\exp\left(-\frac{\Gamma-\Gamma^{(k)}}{2}t_1\right)\right]+\R^{(o)}\right\}.\nonumber
\end{eqnarray}
The efficiencies of the reflection into the CAOM and the AOM are,
according to Eqs.~(\ref{Eff1}) and (\ref{Eff2}), given by
\begin{equation}
\underline\eta_\mathrm{\, ref}^{(k)}=\frac{4\Gamma\Gamma^{(k)}}{(\Gamma+\Gamma^{(k)})^2}
\left|\frac{\T^\mathrm{(o)}\T^\mathrm{(c)}}{\Gamma}+
\R^\mathrm{(o)}\right|^2,\label{EtaCAOM}
\end{equation}
\begin{equation}
\overline\eta_\mathrm{ref}^{(k)}=\left(\frac{\Gamma-\Gamma^{(k)}}{\Gamma+\Gamma^{(k)}}
\right)^2
\left|\frac{2\T^\mathrm{(o)}\T^\mathrm{(c)}}{\Gamma-\Gamma^{(k)}}+
\R^\mathrm{(o)}\right|^2.
\end{equation}

From Eq.~(\ref{EtaTotal}) the total efficiency of the reflection,
$\eta_\mathrm{ref}^{(k)}$, can be found to be
\begin{equation}
\eta_\mathrm{ref}^{(k)}=\frac{D}{\Gamma\left(\Gamma+\Gamma^{(k)}\right)}+
\left|\R^\mathrm{(o)}\right|^2,\label{EtaTotalK}
\end{equation}
where
\begin{equation}
D=4\left|\T^{(o)}\right|^2\left|\T^{(c)}\right|^2+4\Gamma\Re\left(\T^{(o)}\T^{(c)}
R^{(o)\ast}\right).\label{FDefinition}
\end{equation}
This efficiency can be measured, for example, by reflecting the pulse being in
a coherent state and comparing the amplitudes before and after reflection.

\section{Measurement procedure}

For the formulation of the first step of the measurement procedure,
we consider the special case $\Gamma\ll\Gamma^{(k)}\ll\textrm{FSR}$.
In this case, as it follows from Eqs.~(\ref{OutPulse}) and
(\ref{EtaTotalK}) in the limit
$\Gamma^{(k)}/\Gamma\rightarrow+\infty$, the TP does not change its
form during reflection (and partial absorption and scattering)
according to
\begin{equation}
U^\mathrm{out}\!\left(t_1\right)=\R^{(o)}U_0^\mathrm{in}\!
\left(t_1\right),
\end{equation}
\begin{equation}
\eta_\mathrm{ref}=\left|\R^{(o)}\right|^2.
\end{equation}
Hence, one can conclude that such pulses do not
couple with the intracavity field and the efficiency of their
reflection is completely defined by the value of $\R^{(o)}$. This
efficiency as well as the corresponding phase shift can be measured.
According to Eq.~(\ref{constr2}) with such kinds of
experiments one can check the strength of the noise term
$\hat{C}^\mathrm{(o)}(t)$ in the
input-output relation (\ref{IOR}). Therefore, by reflecting an
appropriate pulse by the cavity one can measure the value of
$\R^{(o)}$ and hence of $|\A^{(o)}|$.

The second step of the procedure is the measurement of the two total
efficiencies of the reflection $\eta_\mathrm{ref}^{(1)}$ and
$\eta_\mathrm{ref}^{(2)}$ for TP's defined by Eq.~(\ref{TestPulse})
with two different parameters $\Gamma^{(1)}$ and $\Gamma^{(2)}$,
respectively. One can consider Eq.~(\ref{EtaTotalK}) for $k=1,2$ as
a system of two algebraic equations for the variables $\Gamma$ and
$D$. Resolving it we obtain
\begin{equation}
\Gamma=\frac{\Gamma^{(2)}\left(\eta_\mathrm{ref}^{(2)}-\left|\R^{(o)}\right|^2
\right)-\Gamma^{(1)}\left(\eta_\mathrm{ref}^{(1)}-\left|\R^{(o)}\right|^2
\right)}{\eta_\mathrm{ref}^{(1)}-\eta_\mathrm{ref}^{(2)}},\label{GammaMeasurement}
\end{equation}
\begin{equation}
D=\Gamma\left(\Gamma+\Gamma^{(k)}\right)\left(\eta_\mathrm{ref}^{(k)}-
\left|\R^{(o)}\right|^2\right).\label{FMeasurement}
\end{equation}
Therefore, this part of the measurement procedure allows one to
obtain the values of cavity decay rate $\Gamma$ and of $D$,
Eq.~(\ref{FDefinition}), by reflection of different pulses at the
cavity.

\begin{figure}[htb!]
\includegraphics[clip=, width=0.95\linewidth]{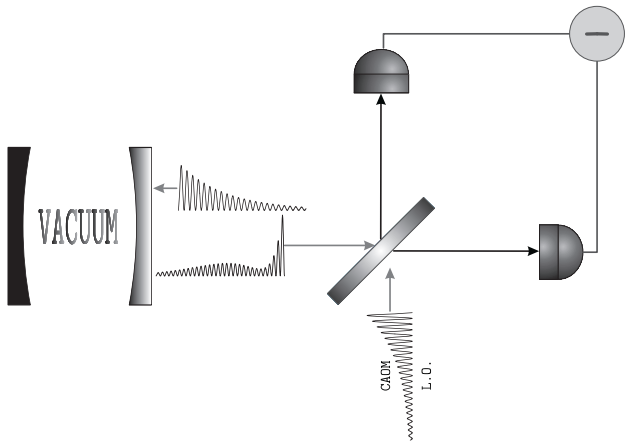}
\caption{\label{fig2} The scheme of measuring the efficiency
$\underline\eta_\mathrm{\, ref}$ of the reflection into the CAOM. The
local-oscillator field in the standard scheme of homodyne detection
is prepared in the form of the CAOM.}
\end{figure}

The third step of the procedure is the measurement of the efficiency
of the reflection into the CAOM as given by Eq.~(\ref{EtaCAOM}).
Because of the specific form of this expression one can consider,
without loss of generality, the efficiency
$\underline\eta_\mathrm{\, ref}$ for the case of
$\Gamma=\Gamma^{(k)}$. This value can be measured by balanced
homodyne detection~\cite{BalancedHomodyning} of the reflected pulse
being in the coherent state. The local oscillator field is prepared
in the form of the CAOM with an arbitrary phase, see
Fig.~\ref{fig2}. Such a procedure allows one to measure only the
quadrature of the CAOM with suppression of information about the
state of the AOM, see discussion of cascaded homodyning in
Ref.~\cite{Semenov}. Comparing the absolute values for the coherent
amplitudes of the input pulse and the CAOM, one can find the
corresponding efficiency $\underline\eta_\mathrm{\, ref}$.

Combining Eq.~(\ref{EtaCAOM}) for $\Gamma=\Gamma^{(k)}$ and
Eq.~(\ref{FDefinition}), one obtains the expression for
$|\T^{(o)}|^2|\T^{(c)}|^2$,
\begin{equation}
|\T^{(o)}|^2|\T^{(c)}|^2=\frac{D}{2}+\Gamma^2\left(\left|\R^{(o)}\right|^2
-\underline\eta_\mathrm{\, ref}\right),\label{Magnitudes}
\end{equation}
and than Eq.~(\ref{FDefinition}) can be used for expressing the phase of
$\T^{(o)}\T^{(c)}$ as
\begin{equation}
\arg\T^{(o)}\T^{(c)}=\arccos\frac{D-4|\T^{(o)}|^2|\T^{(c)}|^2}{4\Gamma|\T^{(o)}||\T^{(c)}|}
-\arg\R^{(o)}.
\end{equation}
Therefore, this stage of the measurement procedure allows one to
obtain the value of $\T^{(o)}\T^{(c)}$.

As the fourth step of the measurement procedure one should separate
the coefficients $\T^{(o)}$ and $\T^{(c)}$. In the most general case
it is impossible to do this by using only the information about
reflection by the cavity. It can be performed directly by measuring
the efficiency of the quantum state extraction given by
Eq.~(\ref{EtaExt}) and the corresponding phase. This kind of
experiment allows one, in principle, to obtain the value of
$\T^{(o)}$ and evaluate $\T^{(c)}$ using the knowledge of
$\T^{(c)}\T^{(o)}$.

As it follows from Eqs.~(\ref{constr2}) and (\ref{constr3}), the
difference between the absolute values of $\T^{(o)}$ and $\T^{(c)}$
is caused by deviation of $\left|\R^{(o)}\right|^2$ from $1$.
According to Eqs.~(\ref{cr2}) and (\ref{constr2}), the additional
noise term $\hat{C}^{(o)}\!\left(t_1\right)$ in the input-output
relation~(\ref{IOR}) differs from $0$ for such cases. This term
describes unwanted losses of the input field inside the coupling
mirror under reflection by the cavity. From the other hand, for
cavities with a negligible value of this noise, the first step of
the measurement procedure reveals that $\left|\R^{(o)}\right|^2=1$,
consequently for this case
\begin{equation}
\left|\T^{(o)}\right|=\left|\T^{(c)}\right|=
\sqrt{\left|\T^{(o)}\right|\left|\T^{(c)}\right|}.
\end{equation}
For such cavities the absolute values of
$\T^{(o)}$ and $\T^{(c)}$ are easily obtained.

\section{Conclusions}

Leaky optical cavities with unwanted noise are characterized by
several parameters. We have demonstrated that they are uniquely
related to the efficiencies and phase shifts of the reflection of
different light pulses. The proposed procedure of measuring the
cavity parameters is based on the determination of these
efficiencies. This can be done by comparing the coherent amplitudes
of the incident and reflected light fields. Among others, it has
been demonstrated that the specific noise term in the input-output
relation can be measured through the reflection of a pulse, whose
spectrum is much wider than the cavity-decay rate but narrower than
the free spectral range. Altogether, the possibility to measure all
the parameters allows one to completely characterize the quantum
noise effects of a given nonideal cavity.

\subsection*{Acknowledgements} This work was supported by Deutsche
Forschungsgemeinschaft. One of the authors (A.A.S.) gratefully
acknowledge the President of Ukraine for financial support. He also
thanks A.V.~Turchin (Institute of Physics, NAS of Ukraine) and
E.E.~Mikhailov (College of William and Mary) for useful comments.

\end{document}